# Multi-Objective Aerodynamic Optimization of External Compression Supersonic Intake using Kriging/MOGA

J. P. S. Sandhu,[1] M. Bhardwaj,[2] N. Ananthkrishnan,[3] and A. Sharma[4]
*Yanxiki Tech, 152 Clover Parkview, Koregaon Park, Pune, 411001, India*

I.-S. Park[5]
*Agency for Defense Development, Daejeon, 305-600, South Korea*

**A formal optimization procedure is reported for an external-compression supersonic intake with the twin objectives of maximizing the total pressure recovery while simultaneously minimizing the intake pressure drag. Prior to that, two key design modifications are introduced to the supersonic intake. First, a small cowl offset distance is applied, deliberately violating the shock-on-lip condition. This ensures an attached external shock at the cowl, which is beneficial, in return for a small, controlled flow spillage. Second, in a novel development, an internal wedge angle is introduced at the cowl lip, which forces the terminal shock to be a strong oblique shock. This also helps anchor the terminal shock at the cowl lip for a range of intake back-pressure values, and reduces the cowl external angle with respect to the free stream. The optimization problem is formulated based on axiomatic design theory with the two design parameters innovatively selected as the Oswatitsch ramp angle triplets and the cowl internal wedge angle, respectively. A multi-objective genetic algorithm (MOGA), assisted by a Kriging meta-model with infilling, is run in tandem with a RANS solver, to iteratively compute the Pareto front. The optimal design yields ramp angles that are somewhat smaller than the theoretical Oswatitsch values, yet returns a total pressure recovery very close to the Oswatitsch optimum while reducing the intake pressure drag significantly.**

---

[1] Senior R&D Engineer, Member AIAA; jatinder@yanxiki.com

[2] R&D Engineer; megha@yanxiki.com

[3] Technical Consultant, Associate Fellow AIAA; akn.korea.19@gmail.com, akn@aero.iitb.ac.in

[4] Principal Engineer, Member AIAA; anurag@yanxiki.com

[5] Principal Researcher, Propulsion Division

## Nomenclature

| | | |
|---|---|---|
| $C_d$ | = | intake pressure drag coefficient |
| $d$ | = | diameter of rounded cowl lip |
| $h$ | = | height of the intake |
| $M$ | = | Mach number |
| $P$ | = | total pressure |
| $p$ | = | static pressure |
| $T$ | = | static temperature |
| $\alpha$ | = | ramp shock angle |
| $\beta$ | = | cowl lip geometric angle |
| $\Delta s$ | = | cowl lip offset distance |
| $\delta$ | = | ramp angle with respect to the reference axis |
| $\delta_e$ | = | cowl lip external wedge angle |
| $\delta_4, \delta_i$ | = | cowl lip internal wedge angle |

*Subscripts*

| | | |
|---|---|---|
| *1,2,3* | = | refers to first, second, third ramp, respectively |
| $\infty$ | = | free stream condition |

## I. Introduction

Air intakes are a critical element in supersonic and hypersonic aircraft and missiles. A well-designed air intake is key to efficient operation of air-breathing engines in high-speed flight over a range of flight conditions and flow incidence angles. The primary purpose of the air intake is to deliver the required air mass flow rate to the engine with as little loss as possible. The loss is usually estimated in terms of the total pressure recovery (TPR), that is, the ratio of total pressure at the engine entry station to the free stream total pressure. In the process of slowing down the supersonic free stream flow to an acceptable Mach number at the engine entry station, the air intake passes the flow through a series of shocks, raising the static pressure and temperature, so that the fuel-air mixture in the combustor may readily ignite and burn. However, the passage of air flow through each shock inevitably results in a loss in total pressure, which in turn negatively impacts the thrust output of the engine. Some total pressure is also lost within the intake duct due to flow separation and viscous losses. Thus, optimizing the supersonic air intake to maximize the total pressure recovery has been an important design objective [1, 2].

Additionally, air intakes are required to deliver as uniform an air mass flow as possible at the engine entry station, which is more critical in the case of gas turbine engines and less so for ramjet combustors. In any case, the flow non-uniformity or distortion at the intake exit station (equivalently, the engine entry station) is a factor of merit for air intake design [3]. Often distortion arises because the air intakes are required to be curved, which may cause regions of flow separation, and even reverse flow, within the intake duct. Stealth designs may require serpentine S-shaped intakes to reduce the radar signature due to engine blades [4, 5]. Many aircraft utilize Y-shaped intake ducts which merge the flow from two side intakes into a single passage before delivering the air mass flow to the engine. In these cases, there is a possibility of unequal air mass flow rates from each side intake resulting in non-uniform flow in the merged duct [6, 7]. Even otherwise, the air mass flow passing through a series of oblique shocks (either wedge-type or conical) will inevitably be turned through a certain angle, and must be turned back through a curved intake duct so as to become parallel to the engine axis. If the duct curvature is too sharp, the likelihood of flow separation is high, with the attendant total pressure loss. On the other hand, a long, gradually curved intake duct presents a larger frontal area to the flow, increasing the pressure drag, which we discuss next.

In the process of slowing down a supersonic flow, any intake duct incurs a drag penalty. The intake ramps, both of the wedge and cone type, present a frontal area to the flow, as does the external surface of the cowl which encloses the intake duct. The higher static pressure due to the series of intake shocks acting on these frontal surfaces can be a major source of drag. In case of the cowl, if a bow shock forms at the cowl lip, the higher static pressure can result in higher drag. Another source of drag is due to flow spillage over the cowl lip, which happens when the entire capture stream-tube does not enter the intake duct. Finally, there is the skin friction drag due to viscous effects on all external intake surfaces. From an aircraft performance point of view, the objective is to maximize the difference between thrust and drag; thus, it is not enough to design a supersonic intake to simply maximize the total pressure recovery, as one needs to minimize the intake drag simultaneously as well.

Yet another factor must be taken into consideration when designing a supersonic intake. The intake shock structure, and hence the total pressure recovery, also depends on the combustor operating pressure (also called the intake back-pressure). At a higher back-pressure, the shock system in the intake duct is usually pushed forward towards the cowl lip, and this results in a higher total pressure recovery. Cases with the terminal shock located within the intake duct are called supercritical. In the best case, the terminal shock is located at the cowl lip, called the critical condition, which gives the highest total pressure recovery of all. Any further increase in the back-pressure can cause the terminal shock to be located ahead of the cowl lip, on the intake ramp, disrupting the ramp shock structure and causing additional flow spillage. This is a subcritical flow condition, called *unstart* if the terminal shock is static or *buzz* if it is oscillatory, and is generally undesirable [8]. The various intake operating conditions are usually captured in the form of an intake characteristic, which is a curve of total pressure recovery versus mass flow ratio (defined as the ratio of the actual mass flow rate of air entering the intake to the maximum possible mass flow rate of air that the intake could theoretically handle) for different intake back-pressure values, as shown in Fig. 1. Ideally, one would like to operate the intake at the critical condition, where the total pressure recovery is the highest. However, due to fluctuations in the combustor pressure, as well as disturbances in the flow, such as due to atmospheric turbulence, stable operation at the critical condition is not possible. Intakes are therefore operated at a safe distance from the critical condition, called the *P4 margin* [9], in a supercritical condition at a lower back-pressure than the critical back-pressure, as marked in Fig. 1. Nevertheless, it is common to optimize the intake design at the critical condition rather than the actual operating supercritical point.

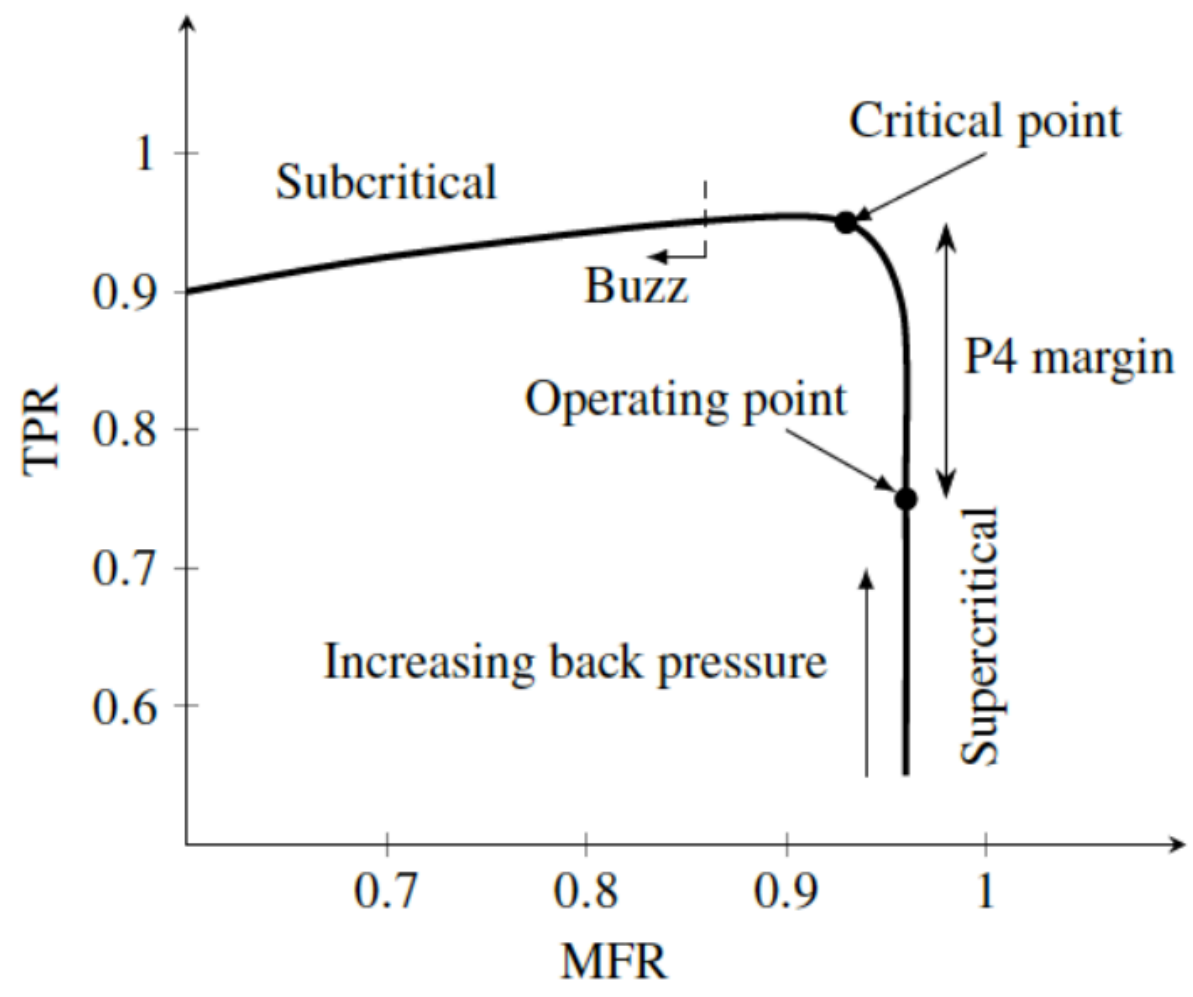

**Fig. 1 Typical intake characteristic of a supersonic intake showing the variation of total pressure recovery (TPR) with mass flow ratio (MFR), parameterized by the intake back-pressure.**

In fact, the optimal arrangement of intake ramp shocks to maximize the total pressure recovery was derived as long back as 1947 and is called the *Oswatitsch criterion* [10]. According to this criterion, in a system of (*n-1*) oblique shocks and one normal shock, the maximum total pressure recovery is obtained when the shocks are of equal strength; that is, when the total pressure recovery across each shock is the same. While this criterion generally holds for both rectangular (wedge-type) and conical (spike-type) intakes, for the case of rectangular intakes, this is equivalent to demanding that the Mach numbers normal to the individual shocks are equal. Using this method, one can deduce the ideal ramp angles and the associated oblique shocks, as well as the terminal normal shock at the cowl lip, in order to obtain the best possible total pressure recovery for a given flight condition, knowing the total number of shocks desired. One can, however, imagine a solution where the intake ramp angles are a little smaller than the Oswatitsch solution, whereby the total pressure recovery is somewhat less than optimal; then the ramp walls and the external cowl surface for the sub-optimal solution would present a smaller frontal area to the flow, and the smaller ramp angles would create weaker shocks, hence lower pressure on these surfaces, thus creating less pressure drag. That is to say, one could trade off some total pressure recovery to obtain lower intake pressure drag. In other words, the Oswatitsch criterion gives the optimal intake design only when maximizing total pressure recovery is the sole objective. If both maximum total pressure recovery and minimum intake pressure drag are simultaneously sought, then one may expect the ramp angles to deviate from the ideal Oswatitsch solution. In fact, there may be a range of solutions offering a trade-off between total pressure recovery and intake pressure drag (called a Pareto front in the language of multi-objective optimization), leaving the designer to choose between them.

The application of modern optimization methods to the intake design problem has attracted much research attention in recent years. Reference [11] presented a methodology, using the Oswatitsch criterion, aimed at optimizing the design of a two-dimensional, mixed compression, two-ramp supersonic intake by maximizing total pressure recovery while matching the engine's mass flow demand. Reference [12] presented an optimization strategy for two-dimensional high-speed missile inlets, which employed a semi-empirical flow solver and a genetic algorithm, integrated within an automated loop, to maximize the total pressure recovery of the intake. The methodology was expanded in Refs. [13, 14] to include three-dimensional intakes. The work in Refs. [15, 16] introduced RANS computation for aerodynamics prediction, and included theoretical engine performance models into the intake optimization process. Reference [17] reports on a design code to perform aerodynamic design and analysis of external compression supersonic intakes, with a focus on determining intake geometry and performance. In other work, Ref. [18] integrated CFD software and the Dakota optimization package for optimization of hypersonic intakes. It can be seen that most researchers [11, 12, 14-16, 18] either explicitly use the Oswatitsch criterion to determine the intake ramp angles and the corresponding shock strengths, or the optimization scheme to maximize the total pressure recovery yields a solution for the intake ramps and shocks that is very close to the Oswatitsch solution. Also, it may be noticed that many authors [12, 14] use analytical or empirical relations to evaluate the intake aerodynamics, which may not be sufficiently accurate to capture the intake flow with boundary layers facing adverse pressure gradients, flow separation and reattachment due to sharp edges or curved walls, and vortex shedding.

In this work, we improve upon the prevalent supersonic intake design procedure in several ways, as described below. Firstly, we approach the supersonic intake design as a multi-objective optimization problem where the total pressure recovery is to be maximized and the intake pressure drag minimized at the same time. As we will show, this yields a family of optimal solutions that have intake ramp angles somewhat smaller than that provided by the Oswatitsch criterion. While the Oswatitsch solution is a good starting point for the optimization, it turns out to not be the best solution when the twin objectives of total pressure recovery and intake drag are considered together. Secondly, we use a formal optimization tool, namely the Multi-Objective Genetic Algorithm (MOGA), to carry out all the optimization calculations for this work. Other authors [12, 14-16, 18] have used formal optimization methods as well, but the use of MOGA allows us to generate a family of optimal solutions which offer a trade-off between maximizing total pressure recovery and minimizing intake drag. Thirdly, all evaluations of the intake aerodynamics provided to the optimization tool are obtained by running a steady Reynolds-averaged Navier-Stokes (RANS) solver; hence, viscous effects and flow separation phenomena in the intake geometry are adequately captured and reflect in the optimization solution. In fact, the optimization tool and the RANS solver operate in tandem: the RANS solver provides an initial set of data points, and after every optimizer iteration, an additional set of data points are demanded, which are then supplied by the RANS solver (a procedure known as *infilling*). This sequence of actions continues until the optimizer converges to a solution. A multi-objective optimization of internal compression inlets for hypersonic flight has recently been presented in Ref. [19] where a similar trade-off between compression efficiency and intake drag has been carried out using an evolutionary algorithm and a RANS solver. Fourthly, the formulation of the optimization problem is carried out using a formal method called the Axiomatic Design Theory [20]. This procedure ensures that the Functional Requirements (in our case, maximizing the total pressure recovery and minimizing the intake pressure drag) are well represented in terms of the Design Parameters (to be defined in the following), making it a well-posed problem. Axiomatic Design Theory has been previously applied to aerospace engineering, specifically the problem of controller design for a ramjet engine, where the intake terminal shock location needs to be regulated, so that a specified total pressure recovery is guaranteed [21]. Fifthly, each intake geometry in the space of Design Parameters is so designed that the cowl-lip is slightly offset behind the point where the ramp shocks intersect. In other words, the shock-on-lip condition, which is usually considered the ideal design condition for supersonic intakes, is deliberately violated by a small amount in our designs. Although this causes a little spillage at the cowl lip slightly affecting the mass flow ratio, the advantage is that a much weaker attached shock forms at the cowl lip instead of a strong bow shock which is to be expected for a cowl lip angle of the order of the last ramp angle in external compression intakes. As a result, the pressure drag over the cowl external surface is considerably reduced. Sixthly, we introduce a small effective wedge angle at the internal surface of the cowl at the lip, of the order of *5–10* deg. In fact, this is the second design parameter for our study, besides the intake ramp angles. This cowl-lip internal wedge angle alters the terminal normal shock at the cowl lip to a strong form of the oblique shock. As is well known, the strong oblique shock gives a slightly better total pressure recovery than the corresponding normal shock, and the downstream Mach number is subsonic for both cases. Further, the terminal oblique shock may be expected to remain anchored at the cowl lip for a range of intake back-pressures, whereas the normal shock may relocate itself within the intake duct for lower values of back-pressure. Of course, given a supersonic flow over a wedge of a certain angle, both weak (more usual) and strong oblique shocks are possible, but for the typical values of back-pressure employed in our study, the strong form of the oblique shock was encountered every time.

Each of these developments is described in greater detail in the following sections of this paper. Section II defines the intake geometry parameters and evaluates the theoretical solution to the optimum intake design problem following the Oswatitsch criterion. Section III introduces two intake design modifications — cowl lip offset and cowl internal wedge angle — with some minor corrections for viscous effects to improve the original theoretical solution. The formulation of the intake optimization problem based on axiomatic design theory and the computational procedure using Kriging/MOGA coupled to a RANS solver is reported in Section IV. Section V is devoted to the results of the multi-objective optimization exercise and a discussion of the outcomes. Concluding remarks are presented in Section VI. An appendix provides details of the grid convergence study carried out prior to initiating the RANS simulations.

## II. Geometry Definition and Theoretical Solution

We consider a three-ramp external compression intake, sketched in Fig. 2, as an illustrative example. The free stream conditions correspond to Mach number 2.5 at an altitude of 10 km and angle of attack of zero. In Fig. 2, Points B, C, D denote the beginning of the three ramps, respectively. The corresponding ramp angles are labeled $\delta_1, \delta_2, \delta_3$, and the oblique shock angles at these ramps are denoted by $\alpha_1, \alpha_2, \alpha_3$. Note that the angles are defined with respect to the horizontal reference axis. The cowl lip is originally placed at Point A'; theoretically, this is the point at which the ramp shocks converge satisfying the shock-on-lip condition. Later, on modifying the design, the cowl lip is pushed back by a distance $\Delta s$ to the point A. The intake height $h$ is marked in Fig. 2 as the vertical distance between Point B and Point A (or A'). The cowl lip internal wedge angle is denoted by $\delta_4$. If $\delta_4$ were zero, the cowl would be at angle $\delta_3$, same as the third ramp angle. Thus, $\delta_4$ is the angle by which the cowl is rotated contour-clockwise about point A', as measured from the third ramp angle. The shoulder at point E is rotated by the same angle so that the intake area variation at the cowl lip remains unchanged. The cowl external angle (with respect to the reference axis) then decreases from $\delta_3$ to $\delta_3 - \delta_4$, thereby also reducing the frontal area presented by the cowl to the free stream flow. Depending on the value of the cowl lip internal wedge angle, the terminal shock occurring at the cowl lip is expected to be either a normal shock or a strong oblique one, provided the wedge angle is small enough to have an attached shock. The cowl lip is modeled with a rounded leading edge ($d = 0.0042h$) and a standard value of *4* deg is used for the angle $\beta$ between the cowl-lip external and internal surfaces, as shown in the inset in Fig. 2. Thus, the free stream flow, on encountering the cowl, sees an external wedge angle at the cowl lip of $\delta_e = \delta_3 - \delta_4 + \beta$, whereas the internal flow coming off the third ramp and entering the duct sees an internal wedge angle of $\delta_i = \delta_4$ at the cowl lip.

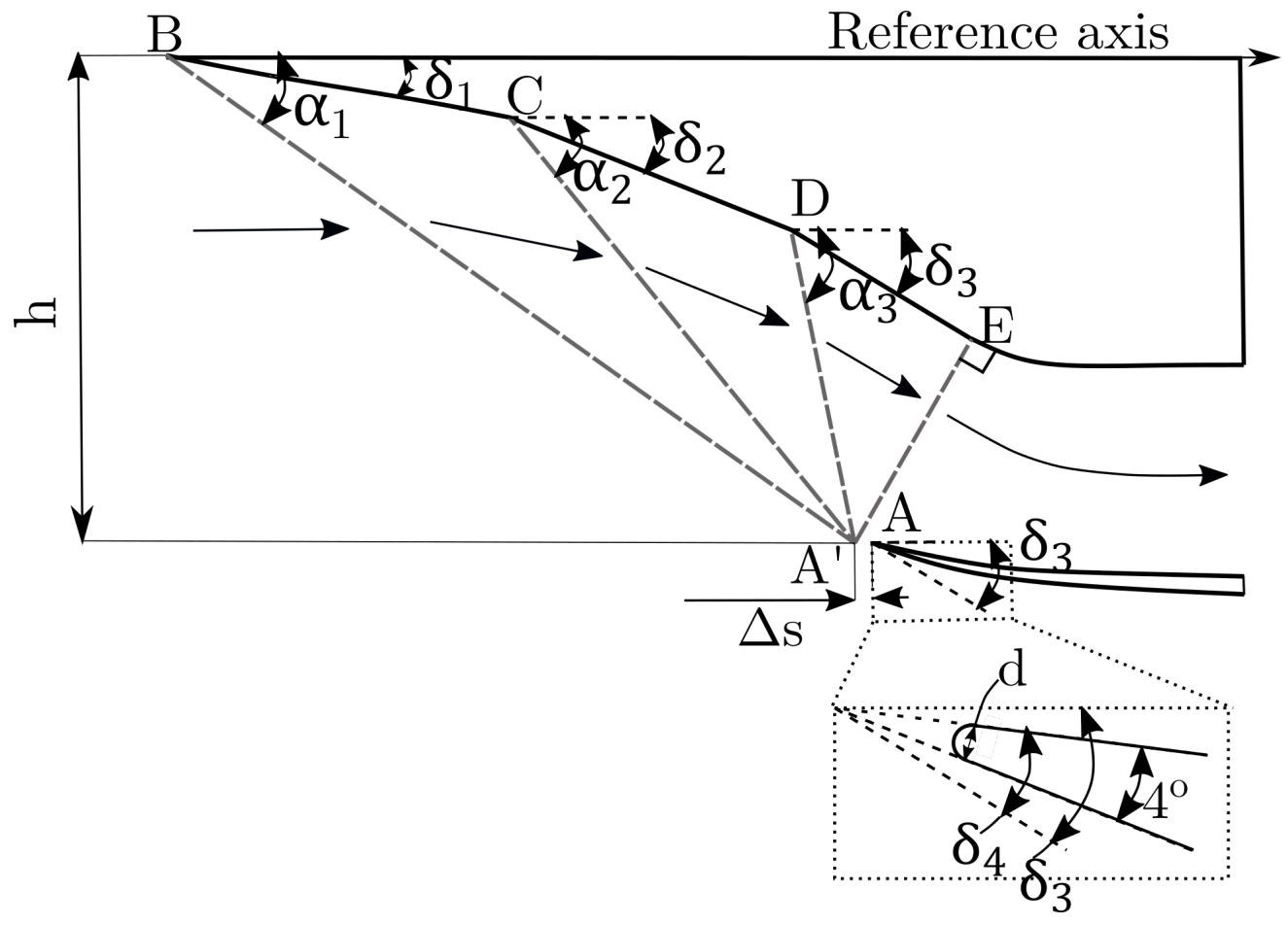

**Fig. 2 Schematic of the three-ramp external compression intake showing the theoretical shock locations, the cowl lip offset $\Delta s$, and the cowl internal wedge angle $\delta_4$.**

The theoretical solution is obtained by applying the Oswatitsch criterion as described in Refs. [11, 22] in terms of the normal component of the Mach number at each shock. The ramp lengths are calculated such that the three oblique shocks and the terminal normal shock are all coincident at the cowl lip (Point A' in Fig. 2); hence, for the theoretical solution, there should be no spillage and the mass flow ratio is expected to be 1.0. The total pressure recovery (TPR) numbers generated in this manner may be presented in the form of TPR contours in the $\delta_1 - \delta_2 - \delta_3$ space. For ease of representation, a two-dimensional cut of these contours is shown in Fig. 3 for a fixed value of $\delta_3 = 33$ deg, which happens to be the optimum angle for the third ramp. From this calculation, the optimum Oswatitsch angles are found to be (*9.7, 20.8, 33.0*) deg, corresponding to the square point in Fig. 3, where the optimum TPR value is found to be *0.922*. Note that this TPR value refers only to the supersonic part of the diffuser. Of interest in Fig. 3 is the line labeled "Oswatitsch Line," which is the locus of points such that each point is geometrically closest (in a Euclidean sense) to the origin from among all the points on a given TPR contour. In other words, for each TPR contour, the point on this line represents the smallest combination of ramp angles for which that TPR value is theoretically possible. Such an Oswatitsch line, defined by ramp angle triplets ($\delta_1, \delta_2, \delta_3$), can be obtained in three-dimensional $\delta_1 - \delta_2 - \delta_3$ space, which we shall use later when formulating the optimization problem.

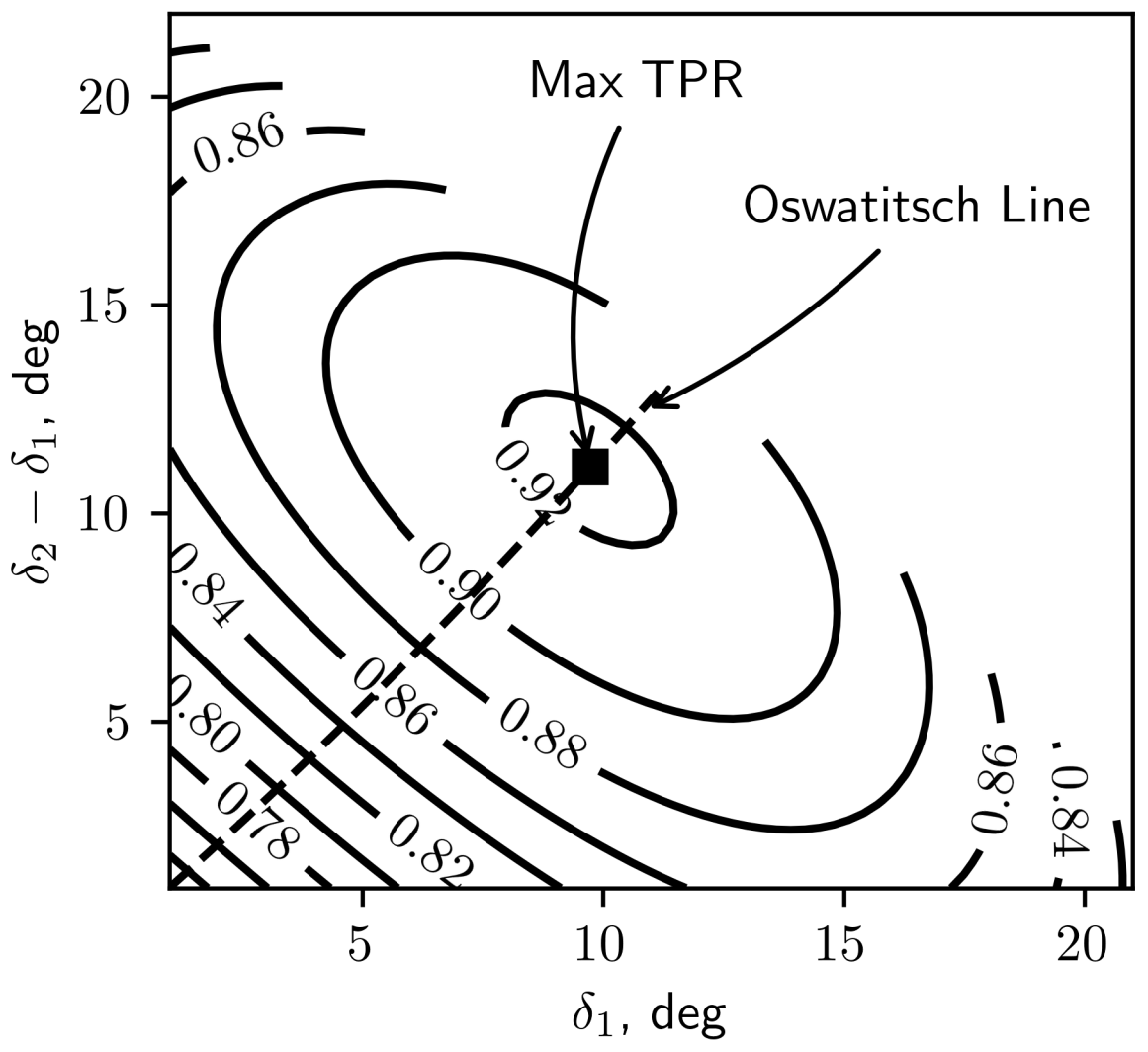

**Fig. 3 Theoretically calculated contours of total pressure ratio in $\delta_1, \delta_2$ space with $\delta_3 = 33$ deg held constant ($\delta_2 - \delta_1$ is plotted on the Y-axis for better visualization).**

### A. Numerical Evaluation

The theoretical optimum design is numerically simulated to evaluate the intake performance parameters, chiefly, the total pressure recovery and the mass flow ratio at this stage, but, later, also the intake pressure drag. The SU2 code [23], a suite of open-source software tools for the numerical solution of Computational Fluid Dynamics (CFD) problems, has been used for this study. The Reynolds-averaged Navier-Stokes (RANS) equations are solved over the domain as outlined in Fig. 4 where a subsonic diffuser has been attached downstream of the three-ramp supersonic intake. The boundary conditions applied at the various surfaces are as marked in Fig. 4. In particular, a back-pressure value is specified at the subsonic outlet boundary at the exit of the subsonic diffuser. The back-pressure ratio (BPR), that is the ratio of the static pressure at the subsonic diffuser exit to the free stream static pressure, is a crucial parameter in determining the intake performance, as indicated in Fig. 1. The AUSM scheme [24] is used for inviscid flux reconstruction, along with the higher-order MUSCL method [25] with Venkatakrishnan limiter [26]. The Euler implicit method is used for time integration. An unstructured grid of approximately *220,000* nodes with *50* viscous layers near the wall is generated using SALOME 9.8. The grid size is based on a grid convergence study, detailed in Appendix A. Menter's Shear Stress Transport (SST) turbulence model [27] is used, and a $Y+$ of less than *1.0* is maintained while generating the meshes since the SST model does not use a wall function. Simulations are taken to have converged when the root-mean-square (RMS) residuals have fallen by at least four orders. The intake total pressure recovery (TPR) is obtained as the ratio of the mass-averaged total pressure at the subsonic diffuser exit plane to the free stream total pressure. The mass flow rate entering the intake duct is evaluated by integrating at a plane slightly behind the terminal shock at the duct entrance as marked by the dashed line in Fig. 4. The intake pressure drag is calculated over the three ramp surfaces and the cowl external surface, which are highlighted in Fig. 4. The drag coefficient is computed with reference to the free stream flow properties and by using a reference area of $1\ m^2$.

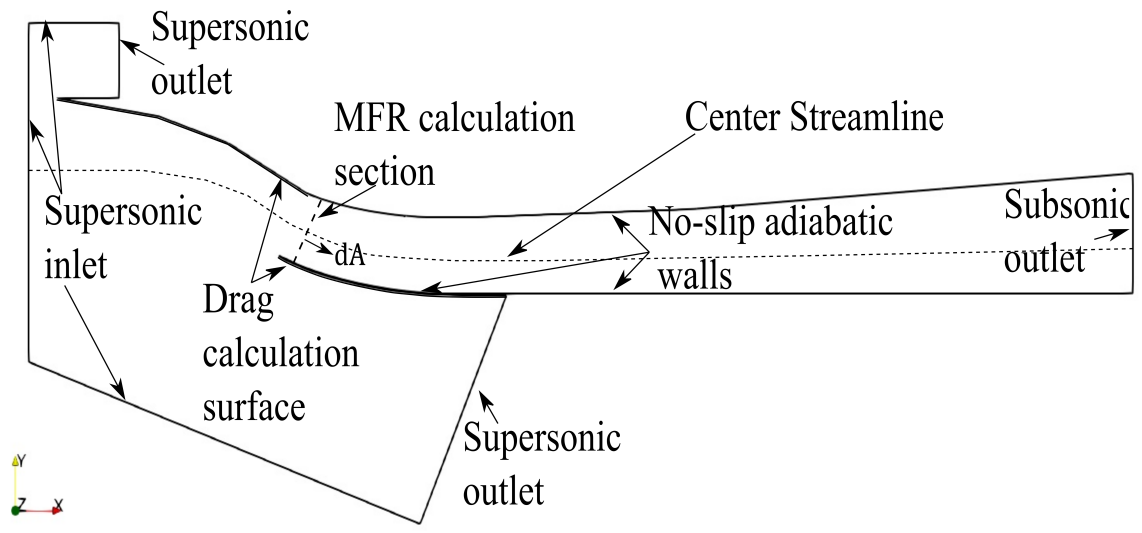

**Fig. 4 Computational domain with boundary conditions and the surfaces for mass flow rate and drag calculation marked.**

Contours of Mach number from the numerical simulation are presented in Fig. 5, where several deficiencies may be noted when compared to the theoretically predicted solution. First of all, the ramp shocks are seen to be displaced from their ideal locations, likely due to viscous effects, so they no longer converge at the cowl lip. The terminal shock is located a little ahead of the cowl lip, marking the operation as nearly critical or perhaps slightly subcritical. The subsonic flow downstream of the terminal shock is seen to spill over the cowl lip, and a significant bow shock has been formed just upstream of the cowl lip. The boundary layer formed over the third ramp is distinctly visible, and there is a region of local supersonic flow within the intake duct, likely due to flow acceleration over the curved shoulder surface. The numerical simulation yields a total pressure recovery (TPR) value of *0.885* for the supersonic diffuser section as against the theoretically predicted optimum value of *0.922*.

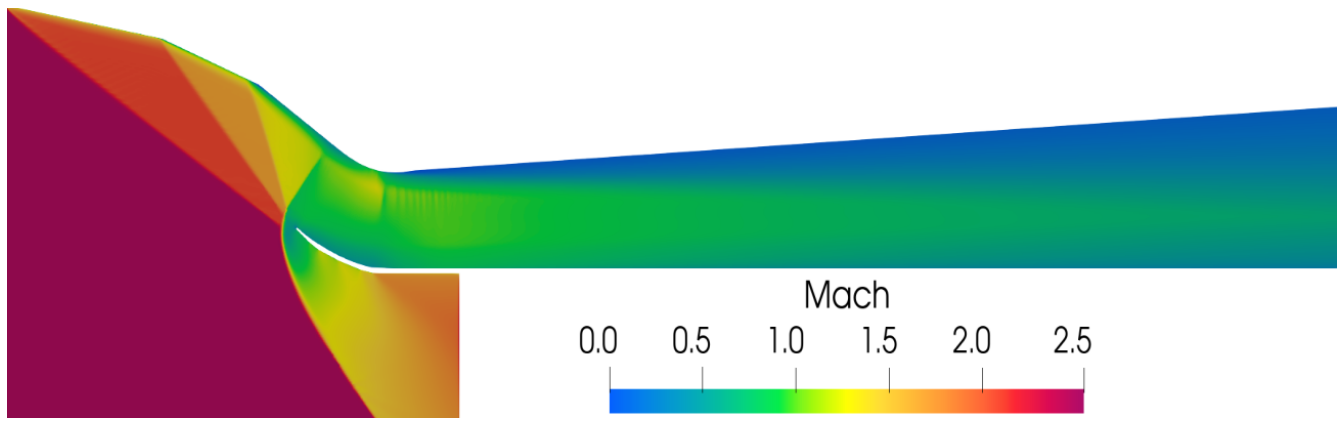

**Fig. 5 Mach number contours for the theoretical intake design at free stream Mach number 2.5, zero angle of attack, and critical back-pressure ratio 12.0.**

Clearly, some design modifications are called for to obviate the deficiencies noted in the numerical solution in Fig. 5 and to recover as much of the theoretically predicted total pressure as possible.

## III. Modified Solution

Two major design modifications, as follows, are carried out to the theoretically designed optimum intake geometry in the previous section.

### A. Cowl Lip Offset

The cowl lip is offset by a fixed distance $\Delta s$ as indicated in Fig. 2 in the axial direction. As explained in Ref. [28], when the shock-on-lip condition is met, it is the free stream flow at zero angle of attack which encounters the cowl external surface. The effective external wedge angle $\delta_e$ at the cowl inevitably causes a detached bow shock to be formed. One way to avoid this is to provide a small cowl offset such that the ramp shocks converge at a point slightly ahead of the cowl lip. In that case, it is (a small part of) the flow over the third ramp that passes over the cowl lip. The effective flow deflection angle at the cowl external surface for this spillage flow is much smaller, of the order of $\delta_e - \delta_3$, and hence an attached external shock is very likely at the cowl lip, which could significantly reduce the pressure drag on the external cowl surface, besides regulating the flow spillage at the cowl lip to a limited value. A fixed value of $\Delta s = 0.0084h$ is maintained for all simulations.

### B. Cowl Internal Wedge Angle

In a novel development, a small cowl internal wedge angle $\delta_4$ is introduced, as described in Fig. 2. An attached strong oblique shock can then be formed at the cowl lip provided the angle $\delta_4$ is less than the limiting value for the shock to remain attached. For the Mach numbers in question, the upper limit on $\delta_4$ is of the order of *10* deg, so an initial guess value of *5* deg is assumed for $\delta_4$, which will be refined further during the optimization process.

### C. Revised Numerical Simulation

Two other minor corrections are carried out before freezing the modified intake geometry. A correction for the viscous boundary layer is applied, similar to the approach in Ref. [29], which results in the angles of the first and second ramps being reduced slightly from their Oswatitsch optimal values. This is to ensure that the ramp shocks do converge to a single point, which is now designed to be $\Delta s = 0.0084h$ ahead of the offset cowl lip, Point A in Fig. 2. The original ramp angles of (*9.7, 20.8, 33.0*) deg are now replaced with the new set of values: (*9.5, 20.7, 33.0*) deg. Secondly, the shape of the subsonic diffuser in Fig. 4 has been altered to allow for a more gradual change in slope in the throat region. While this has little impact on the shock structure upstream in the supersonic part of the diffuser, it may help avoid the regions of local supersonic flow observed within the duct in Fig. 5 and generally reduce the extent of separated flow in the subsonic diffuser duct. The original computational domain and the revised one are sketched in Fig. 6 for comparison. Note that both the cowl lip offset and the cowl internal wedge angle are incorporated in the new domain.

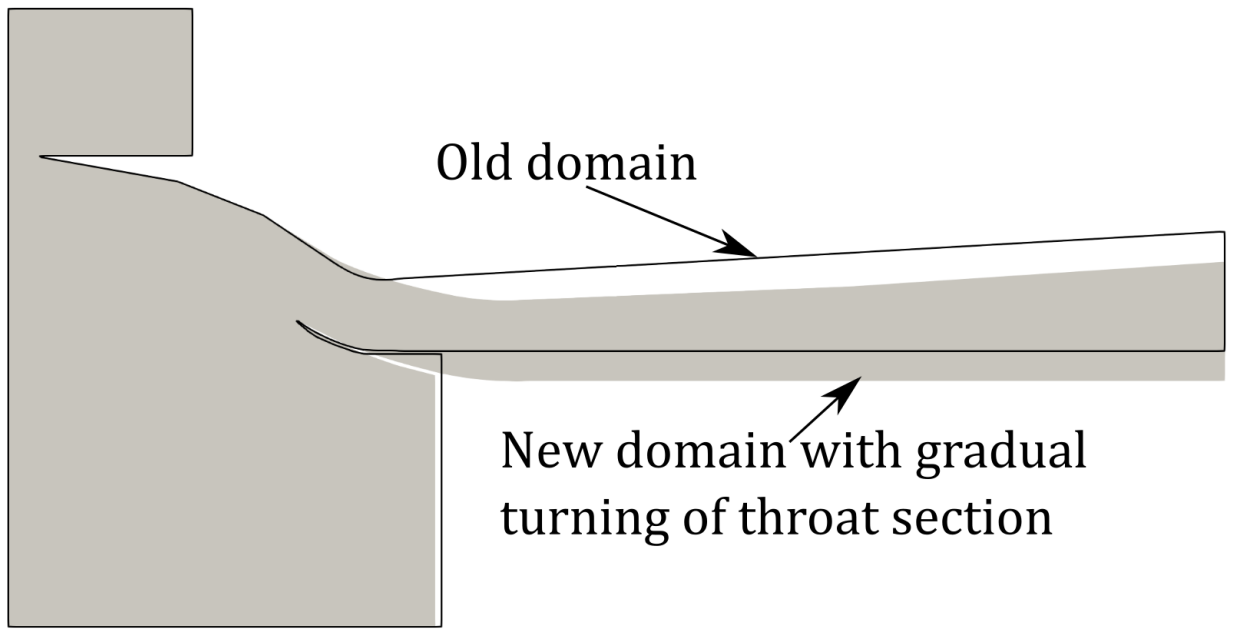

**Fig. 6 Computational domain for the modified intake design at free stream Mach number 2.5, zero angle of attack, and back-pressure ratio 12.0.**

The numerical procedure is identical to that used previously for the original theoretical intake geometry. Mach number contours for the modified intake design are shown in Fig. 7 for the same free stream and back-pressure conditions. Several positive features may be noted from the plot in Fig. 7. The ramp shocks, corrected for viscous effects as stated above, are now seen to converge to a point. While not apparent at the scale of the plot in Fig. 7, this point is slightly upstream of the cowl lip, allowing for a controlled flow spillage at the cowl lip. Consequently, the bow shock at the cowl has been replaced by an attached (nearly attached, since it is offset from the cowl lip by $\Delta s$) shock. The boundary layer over the third ramp has been reduced when compared with Fig. 5, and the regions of local supersonic flow in the throat region of the duct have disappeared. Interestingly, the terminal shock at the entrance to the duct is now a strong oblique shock — this can be verified by inspecting the Mach number downstream of this shock, which is of the order of *0.9*. Had the terminal shock been a normal one, the downstream Mach number would have been around *0.8* or a little lower. Overall, the flow features for the modified design in Fig. 7 appear definitely superior to those in Fig. 5 for the original theoretical geometry.

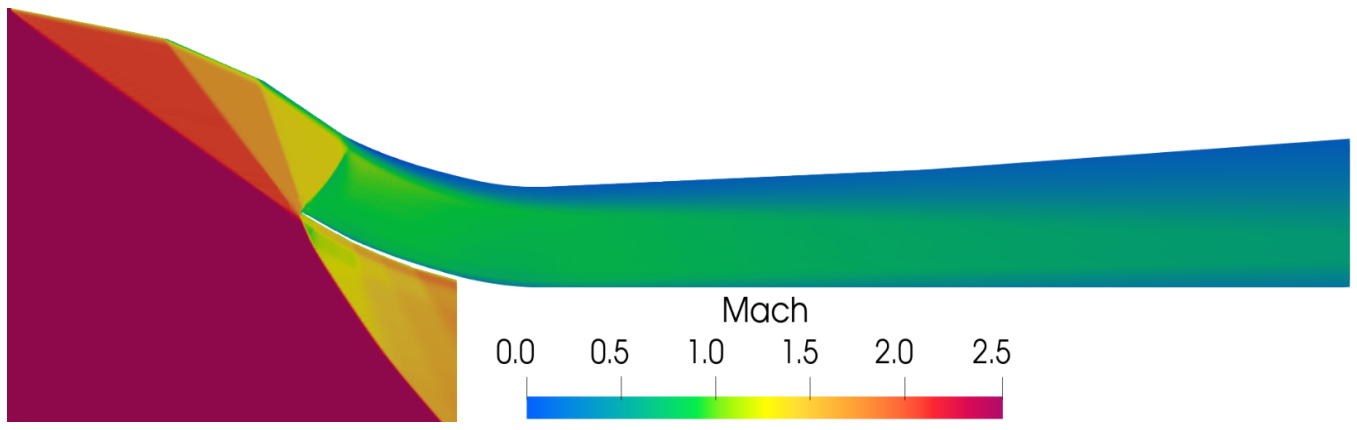

**Fig. 7 Mach number contours for the modified intake design at free stream Mach number 2.5, zero angle of attack, and back-pressure ratio 12.0.**

However, for the simulation in Fig. 7, we have used the same back-pressure ratio of *12.0* as for the original simulation in Fig. 5. With the modified geometry, the critical back-pressure ratio may have changed, and the new critical back-pressure ratio is determined to be *12.4*. Results from a numerical simulation of the modified geometry with *BPR=12.4* are displayed in Fig. 8. In this instance, contours of the density gradient magnitude have been plotted, clearly showing the density changes across the three ramp shocks and the cowl shock, as well as the shear layer originating from the expansion at the shoulder. Notably, the location of the terminal strong oblique shock at the cowl lip in this case is virtually identical to that in Fig. 7, which was for a lower back-pressure ratio. This is suggestive of the fact that the presence of the internal wedge at the cowl lip works to anchor the terminal shock at that location, at least for some range of values of the back-pressure.

Comparing the performance of the original theoretical intake geometry in Fig. 5 with that of the modified design in Fig. 8, at their respective critical values of back-pressure ratio, reveals that the total pressure ratio of the supersonic diffuser part increased from *0.885* to *0.907* for the modified design. The mass flow ratio also improved from *0.955* to *0.983*, thanks to the controlled spillage. Thus, the modifications introduced in this section can certainly be credited with having improved the key intake performance parameters.

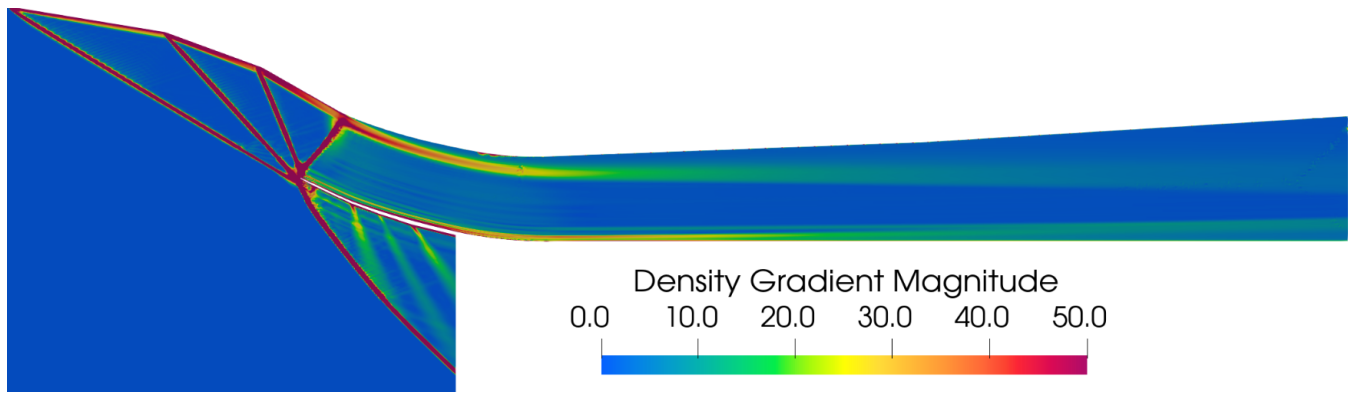

**Fig. 8 Density gradient magnitude contours for the modified intake design at free stream Mach number 2.5, zero angle of attack, and critical back-pressure ratio 12.4.**

# IV. Optimization Problem Formulation

Thus far, the focus has been on optimizing the intake for total pressure recovery (particularly, of the supersonic diffuser), with a secondary interest in maintaining a healthy mass flow ratio. However, in practice, both maximizing the total pressure recovery and simultaneously minimizing the intake pressure drag is of interest. Hence, it is necessary to formulate the intake design exercise as a multi-objective optimization problem. We shall employ the Axiomatic Design Theory [20] to aid in the formulation.

## A. Axiomatic Design Theory

Axiomatic design theory is concerned with the functional requirements (FR), design parameters (DP), and the mapping between the FR and DP from functional space to physical space (the design itself). There are only two axioms, which may be stated as below:

**Axiom 1** [The Independence Axiom] Maintain the independence of the functional requirements.

**Axiom 2** [The Information Axiom] Minimize the information content of the design.

The functional requirements (also called objective functions) in our case, (maximize) the total pressure recovery and (minimize) the intake pressure drag, are clearly independent of each other, hence the first axiom is satisfied. Further, it is desirable that the design is uncoupled in the sense that each design parameter affects only one functional requirement. However, in case an uncoupled design is not possible, independence of the functional requirements must be ensured by decoupling. This is stated in the Corollary below:

**Corollary** [Decoupling of Coupled Design] Decouple or separate parts or aspects of a solution if the functional requirements are coupled or become interdependent in the design proposed.

For the second axiom, functional requirements that are not independent or whose precise value does not matter except that they need to lie within a specified range may be classified as constraints instead. Following this advice, the mass flow ratio is not included as a functional requirement in the optimization process.

One outcome of following these prescriptions is that there can be only as many design parameters as there are functional requirements, namely two, whereas in our problem there are as many as four unknown parameters — the three ramp angles, with the fourth being the cowl internal wedge angle. In an innovative twist, this dilemma can be overcome by defining the triplet of Oswatitsch ramp angles from Fig. 3 to be the first design parameter. These triplets, as we have seen, lie along the Oswatitsch line in Fig. 3 and represent the theoretical optimum (in the sense of being the smallest) values of the ramp angles for a given value of total pressure ratio. Knowing any one ramp angle among these triplets, the values of the other two can be read off from a set of pre-calculated solutions for the Oswatitsch line. The first ramp angle $\delta_1$ is chosen as the representative of a triplet and is designated as the first design parameter. The second design parameter then is the cowl internal wedge angle $\delta_4$. In that case, the mapping between the functional requirements and the design parameters can be stated as:

$$\begin{bmatrix} TPR \\ C_d \end{bmatrix} = \begin{bmatrix} \checkmark & \epsilon \\ \checkmark & \checkmark \end{bmatrix} \begin{bmatrix} \delta_1 \\ \delta_4 \end{bmatrix} \tag{1}$$

The tick mark entries in the design matrix in Eq. (1) indicate a significant dependence of that functional requirement on the corresponding design parameter, whereas $\epsilon$ denotes a weak dependence. The drag coefficient is affected by the ramp angles, as well as the cowl wedge internal angle $\delta_4$ since the strength of the cowl external shock depends on the angle $\delta_e = \delta_3 - \delta_4 + \beta$. In case of TPR, the ramp angles certainly have a strong impact, while $\delta_4$ has a mild influence depending on whether it forms a strong oblique shock or a normal one. Equation (1) does not strictly conform to the requirements of Axiomatic Design Theory due to the presence of the small element $\epsilon$ in the design matrix; however, it may be close enough to the ideal situation to be applicable.

## B. Computational Procedure

Optimization is carried out using Dakota, a software toolkit for optimization and uncertainty quantification of complex systems [30]. The optimization procedure itself is fairly standard, hence it is described only in brief. The design parameter space $\delta_1 - \delta_4$ is initially populated using Latin Hypercube Sampling (LHS) [31]. For each point in this population set, RANS simulation is carried out following the procedure described earlier. In order to compare apples with apples, we need to ensure that the simulation at each combination of $(\delta_1, \delta_4)$ uses the critical value of back-pressure for that intake geometry. This is achieved by running a series of simulations for each point with increasing values of back-pressure, identifying the critical condition from amongst these solutions, and extracting the values of the objective functions (namely, TPR and $C_d$) at the critical state. This sequence of actions is automated. Using the values at these select points, a surrogate model is constructed using the Kriging method to approximate the response of the objective functions elsewhere in the design parameter space [32]. In general, the Kriging fit yields a mean value (also called the trend function) and a measure of the uncertainty (covariance function) at each point. In this work, the universal Kriging option is employed, which uses a polynomial variation for the trend function, along with a Gaussian covariance model. For each of the points in the initial sample, the Kriging mean value is exactly equal to the RANS-derived data value, and there is zero uncertainty at these points. This is because the Kriging mean precisely passes through the given data points, which is different from a least-squares curve fit, for example, which may not pass through any or all of the data points. On the flip side, when there are too many data points, in trying to pass the mean through each one of them, the quality of the approximation at un-sampled points may degrade, that is, the uncertainty or variance in those regions may increase, spoiling the overall fit. This is a problem known as *overfitting* which may be avoided by carefully controlling the number of points included in the sample set.

The response surface obtained by Kriging is fed to the optimizer, which, in our case, is a global multi-objective genetic algorithm (MOGA) [33] provided by Dakota. Using MOGA with Kriging is recommended as it can significantly improve optimization efficiency and accuracy, particularly in cases where there are multiple competing objectives that need to be optimized [34]. The optimizer generates a set of optimal solutions called a *Pareto front*. Among the points on this set, one can trade off TPR for $C_d$, that is, if one point has a better TPR, another point on the Pareto front with a lower TPR will have a better value of $C_d$. This sequence of sampling the design parameter space, generating objective function data by RANS simulation, fitting a response surface by Kriging, and computing the Pareto front by the MOGA optimizer, can be performed iteratively until a converged solution is obtained for the Pareto front.

After each iteration, additional points are added to the design parameter space to improve the approximation of the objective functions by the surrogate model; this procedure is called *infilling* [35,36]. The demand for infill points can come from two sources: from the Kriging algorithm, in regions of high variance in order to reduce the uncertainty in the fit, and, secondly, from the MOGA optimizer, in regions where additional points on the Pareto front are likely. Only a fraction of the infill points suggested at every iteration are added to the sample set in order to reduce the risk of overfitting; this step requires manual intervention. When all the newly suggested infill points are dominated by one or more points on the Pareto front — that is, at least one point on the Pareto front has a better value of both the objective functions — the Pareto front may be considered to have stopped evolving, and further infilling is likely to be futile.

## V. Results and Discussion

The procedure outlined in the previous section is now applied to the optimization of the three-ramp external compression intake under discussion. However, it is instructive to first sketch the likely contours of the Kriging fit in this case, and guess the direction in which the optimization algorithm may lead us. Figure 9 shows the likely contour lines for the two objective functions in the space of design parameters. The initial guess solution based on the Oswatitsch criterion is marked by the filled circle. For the supersonic diffuser TPR, since it is a weak function of the cowl wedge angle, the contours should be nearly vertical lines, with the maximum TPR at the Oswatitsch ramp angle, and falling off to both left and right. This is indicated in the left-hand plot in Fig. 9. For the intake $C_d$ on the right-hand plot, the lowest values are likely to be for small ramp angles and large cowl wedge angles. The contours then appear slanted as shown, with $C_d$ increasing from North-west to South-east on the plot. Starting from the initial guess (filled circle in Fig. 9), clearly the values of both objective functions worsen as one moves to the East and South-east. Moving West, North-west, we lose on TPR but gain on $C_d$. Thus, a trade off between TPR and $C_d$ is likely if we seek solutions roughly North-west of the initial guess point.

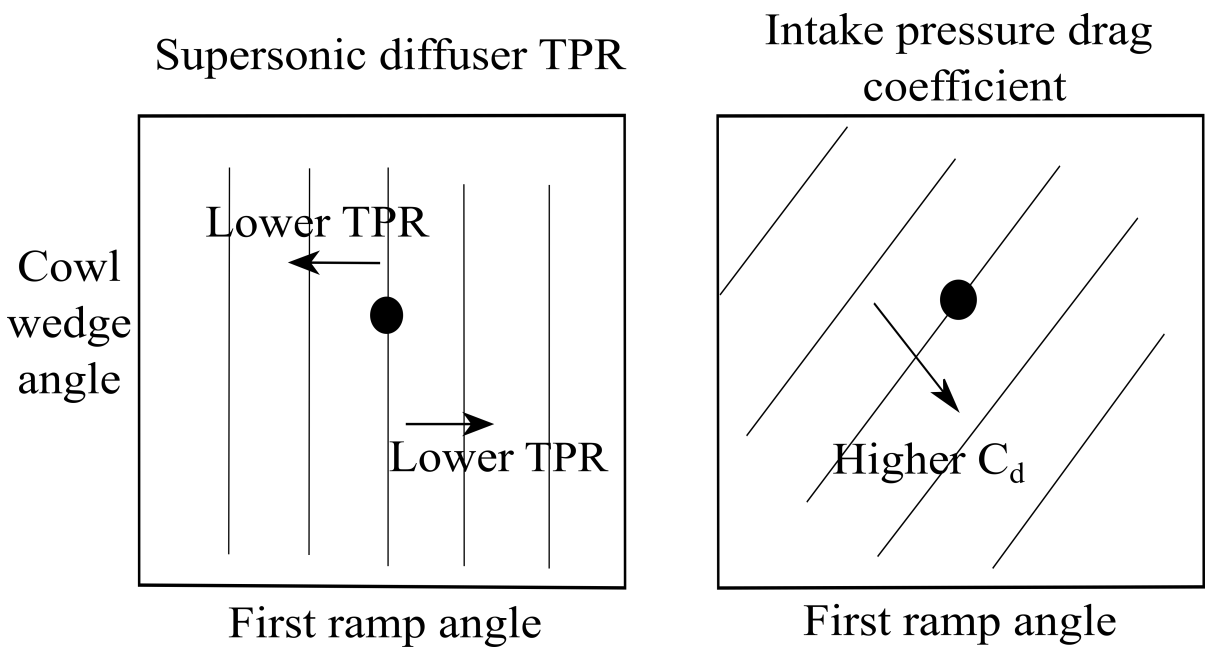

**Fig. 9 Sketch of likely contours from the Kriging fit for the multi-objective intake optimization problem.**

Initially, a sample set consisting of *20* points is chosen based on LHS. After each iteration, additional infill points are included, as discussed previously, and by the third iteration the population set comprises of a total of *31* points. Contours of the two objective functions from a Kriging fit, based on RANS simulations at these *31* points, after three iterations[6] of the optimization procedure are shown in Fig. 10. Note how the general trend is broadly similar to what was anticipated in the sketch of Fig. 9. Further infill points suggested at the end of the third iteration are marked in Fig. 10 by the closed circles. Before proceeding further, one should examine the Pareto front generated at the end of the third iteration; Fig. 11 shows the Pareto front in terms of the two objective functions, plotted as Total

[6] Mean and variance plots from the Kriging fit after each iteration are reported in Ref. [37].

Pressure Loss (TPL=1-TPR) and $C_d$. This is done so that both the objective functions are now being minimized, which is standard practice. The non-dominated points among the *31* sample points where RANS simulations were run (marked by filled squares) have been connected by lines such that a bounded curve is formed in Fig. 11, which represents the Pareto front. Notice that all the infill points suggested at the third iteration (marked by diamonds in Fig. 11) lie either on the Pareto front or well within it, hence there is no need to infill them to the sample data set as they are unlikely to be non-dominated. In that case, the solution after the third iteration may be taken to have converged, and the Pareto front in Fig. 11 has the final evolved form. The Utopia point is marked by a triangle, and the nearest point on the Pareto front to the Utopia point (determined by taking the Euclidean distance) may be assumed to be the optimal solution, marked by a filled circle.

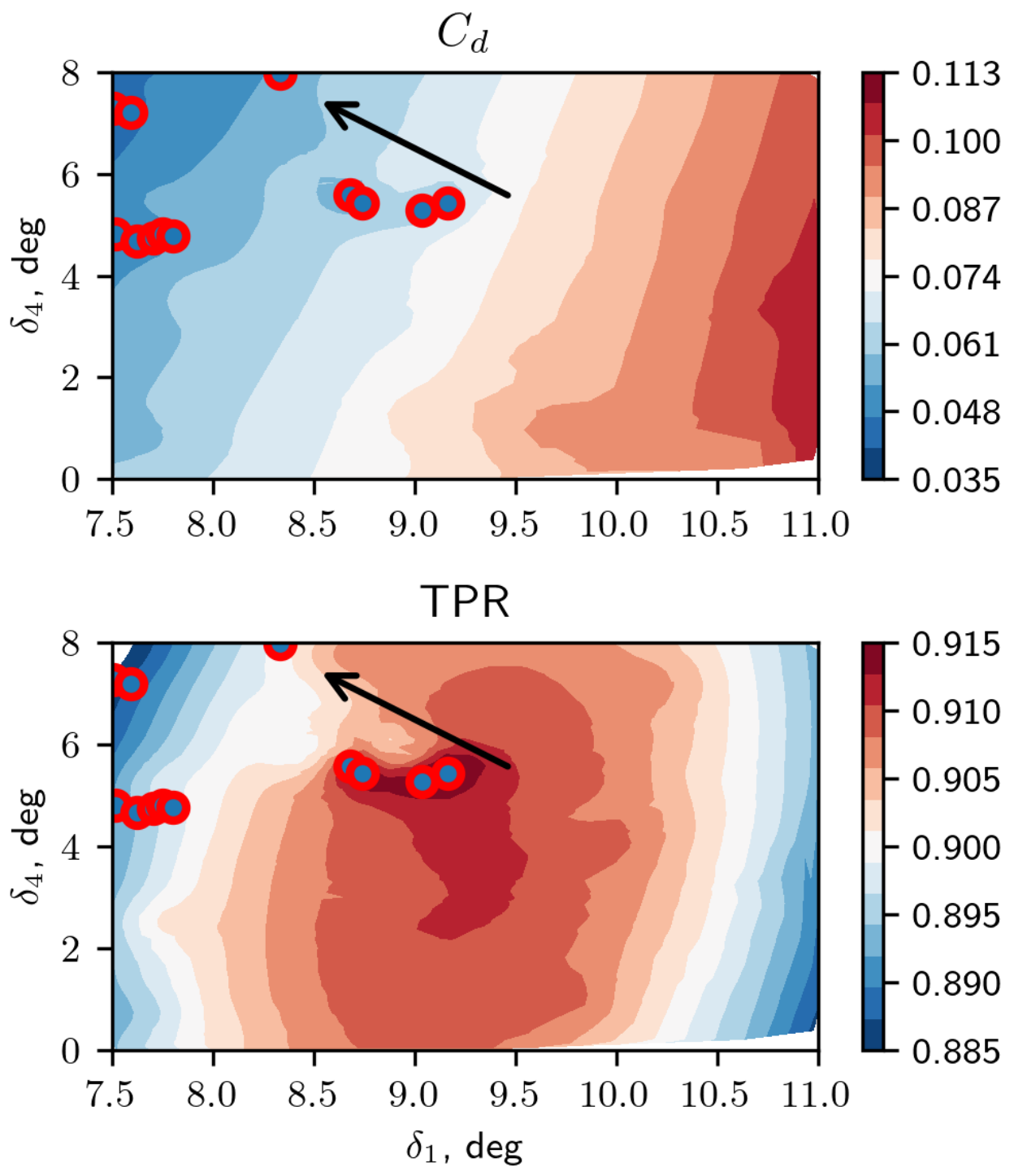

**Fig. 10 Kriging fit for the multi-objective intake optimization problem after third iteration (mean values).**

In the space of objective functions, as plotted in Fig. 11, the bold arrow indicates the shift from the initial theoretical solution (tail of the arrow) to the final optimum solution (tip of the arrow). Compared to the theoretical design, the optimum intake has a slightly larger TPL (that is, a slightly lower total pressure recovery) with a significantly better value of the intake drag coefficient. In other words, the optimizer has traded off a small amount of TPR in return for a notable decrease in $C_d$. The same shift, in design parameter space, is indicated by the bold arrows in Fig. 10. The first ramp angle has been reduced from the Oswatitsch value of *9.5* deg by about *1* deg, while the cowl internal wedge angle has changed from the initial guess value of *5* deg to almost *7.5* deg. Thus, it can be concluded that in case of a multi-objective optimization of an external compression intake, allowing for a trade-off between the total pressure recovery and the intake pressure drag could result in optimum values of the intake ramp angles that are somewhat smaller than the theoretically predicted Oswatitsch angles.

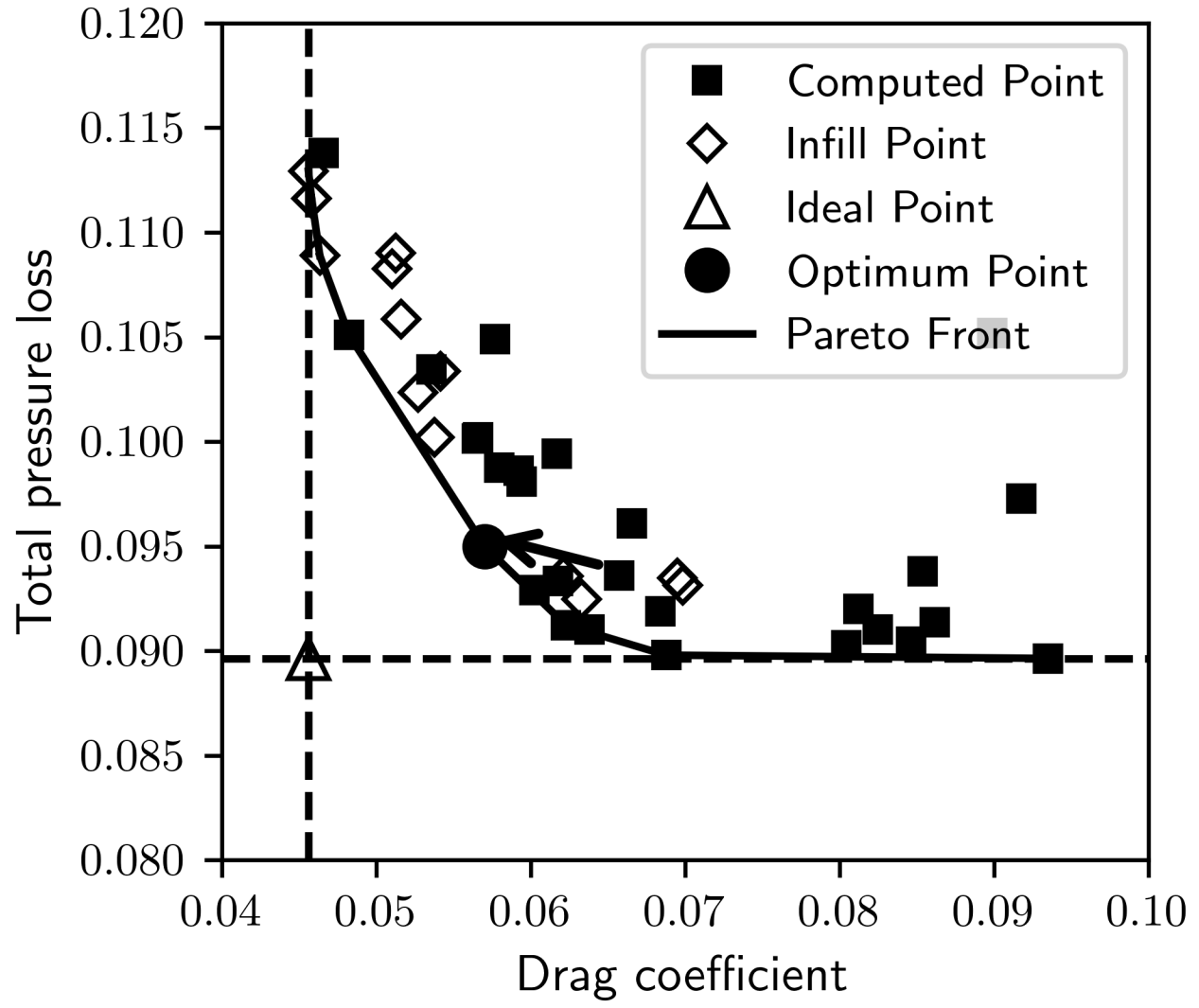

**Fig. 11 Pareto front for the multi-objective intake optimization problem after third iteration.**

A comparative analysis of the intake design and performance parameters for the three intake solutions evaluated in this work — theoretical, modified, and optimal — is presented in Table 1. All the results are from RANS simulations at their respective critical back-pressure ratios. For the theoretical design, the solution is as shown in Fig. 5; for the modified intake, the simulation at critical back-pressure is the one shown in Fig. 8. For the optimal solution, although the quantitative parameters are different, the simulation results are virtually indistinguishable from the contour plot for the modified design in Fig. 8, hence they are not plotted separately. It can be seen from Table 1 that the individual ramp angles for the optimal design are about *1-2* deg smaller than the theoretical Oswatitsch angles. However, the reduction in TPR due to this is quite marginal. On the other hand, there is a marked decrease in the intake drag coefficient from *0.065* to *0.057* between the modified design based on the Oswatitsch angles (adjusted for viscous effects) and the optimum design. This decrease is likely due to the smaller ramp angles and the larger $\delta_4$ for the optimum design, both of which imply a smaller $\delta_e$. For both the modified design and the optimum one, the mass flow ratio is maintained by specifying the cowl lip offset distance (same for both cases). All in all, the trade off between TPR and $C_d$ for the optimum design seems to be a beneficial one.

**Table 1 Comparative analysis of the three intake designs**

| Parameter | Theoretical design (CFD) | Modified design (CFD) | Optimum design (CFD) |
|---|---|---|---|
| $\delta_1, \delta_2, \delta_3$ | 9.7, 20.8, 33.0 deg | 9.5, 20.7, 33.0 deg | 8.5, 18.1, 28.6 deg |
| $\delta_4$ | 0 | 5 deg | 7.5 deg |
| TPR (supersonic diffuser) | 0.885 | 0.907 | 0.905 |
| TPR (subsonic diffuser) | 0.936 | 0.959 | 0.959 |
| Net TPR | 0.828 | 0.870 | 0.868 |
| Mass flow ratio | 0.955 | 0.983 | 0.983 |
| $C_d$ | — | 0.065 | 0.057 |
| Critical BPR | 12.0 | 12.40 | 12.35 |

## VI. Conclusion

Our study has shown that by considering the twin objective functions of maximizing total pressure recovery and simultaneously minimizing intake pressure drag, a range of optimal solutions, represented by a Pareto front, may be obtained for the supersonic intake optimization problem. Interestingly, the best among these solutions (in the sense of being closest in the objective function space to the Utopia value) gives a total pressure recovery (TPR) very close to the Oswatitsch value, and a considerably lower value for the intake drag, for intake ramp angles somewhat smaller than that given by the Oswatitsch criterion. Thus, it follows that intake ramp angles selected based on the Oswatitsch criterion alone for maximizing the TPR may not be the ideal solution in practice. The reduction in intake pressure drag arises from several sources: the smaller ramp angles contribute a little to the lower drag; however, the bulk of the drag reduction comes from two imaginative design features. The first is a slight axial offset in the cowl lip behind the point where the intake ramp shocks meet, thereby deliberately violating the shock-on-lip condition slightly. This ensures an attached external shock at the cowl lip, which produces less drag than a detached bow shock. The flip side of this is a small increase in flow spillage, hence a slight reduction in intake mass flow ratio (MFR). The second feature is to rotate the cowl about the cowl lip point in such a manner as to create an effective internal wedge angle at the cowl lip. This angle needs to be small so that the internal cowl lip shock remains attached, however, it ensures that a strong form of the oblique shock (rather than a normal shock) is created at the cowl lip, with the flow behind the strong oblique shock being subsonic. The wedge angle further weakens the external cowl shock thereby reducing pressure drag on the cowl external surface; at the same time, the strong oblique shock gives a slightly better TPR than the normal shock that would have formed had the cowl wedge angle been zero.

Clearly, the various flow phenomena described above require an accurate RANS solver to evaluate the objective functions being fed to the optimizer for different values of the design parameters. The optimizer itself is able to converge to a solution in three iterations after which the Pareto front stops evolving. This required only a limited number (about *30*) of evaluations of the RANS solver for various combinations of the two design parameters. This was possible due to two devices: One, the three ramp angles were bundled into a triplet (which reduced the space of design parameters from four to two dimensions), and their initial guess value was selected using the Oswatitsch criterion. While, as shown above, the Oswatitsch criterion no longer gives the optimal ramp angles for the multi-objective optimization problem, the optimal solution is not very far from the Oswatitsch values, which aids in quicker convergence of the optimizer. Second, the optimizer suggests a mix of infill points where additional RANS evaluations are sought, some near the expected region of the optimum, others in regions where the variance of the Kriging meta-model is poor, that is, where the data is insufficient or unreliable. Picking too many points of the second kind can lead to overfitting by the Kriging meta-model at the next iteration, thereby damaging the approximation. Instead, selectively picking a few infill points among the suggestions at each iteration means that additional RANS runs can be limited and fewer iterations are required for convergence to be achieved.

## Appendix A: Grid Convergence

A grid convergence study is undertaken to determine the optimal grid resolution for the computational analysis. Three different grid levels are considered, namely, Coarse (*L0*), Medium (*L1*), and Fine (*L2*), with details provided in Table 2. To ensure appropriate resolution of the boundary layer, near-wall viscous layers are incorporated into the meshes. A uniform distribution of *50* viscous layers with a growth ratio of *1.15* is adopted for all the grids. The Mach number along a streamline passing through the center of the capture stream-tube originating upstream of the first ramp and passing wholly into the intake duct is plotted in Fig. 12 to compare the results of the different grid levels. The predictions of *L1* and *L2* meshes are in good agreement, in contrast to the coarse *L0* grid that predicts the terminal shock incorrectly at a slightly upstream location. Based on this observation, the *L1* grid is deemed adequate for all simulations in this study.

**Table 2 Mesh details for grid convergence study**

| Mesh type | Minimum element size (mm) | Maximum element size (mm) | Number of mesh elements |
|---|---|---|---|
| Coarse (*L0*) | 0.0016 | 0.0020 | 79,227 |
| Medium (*L1*) | 0.0008 | 0.0010 | 221,683 |
| Fine (*L2*) | 0.0004 | 0.0005 | 514,691 |

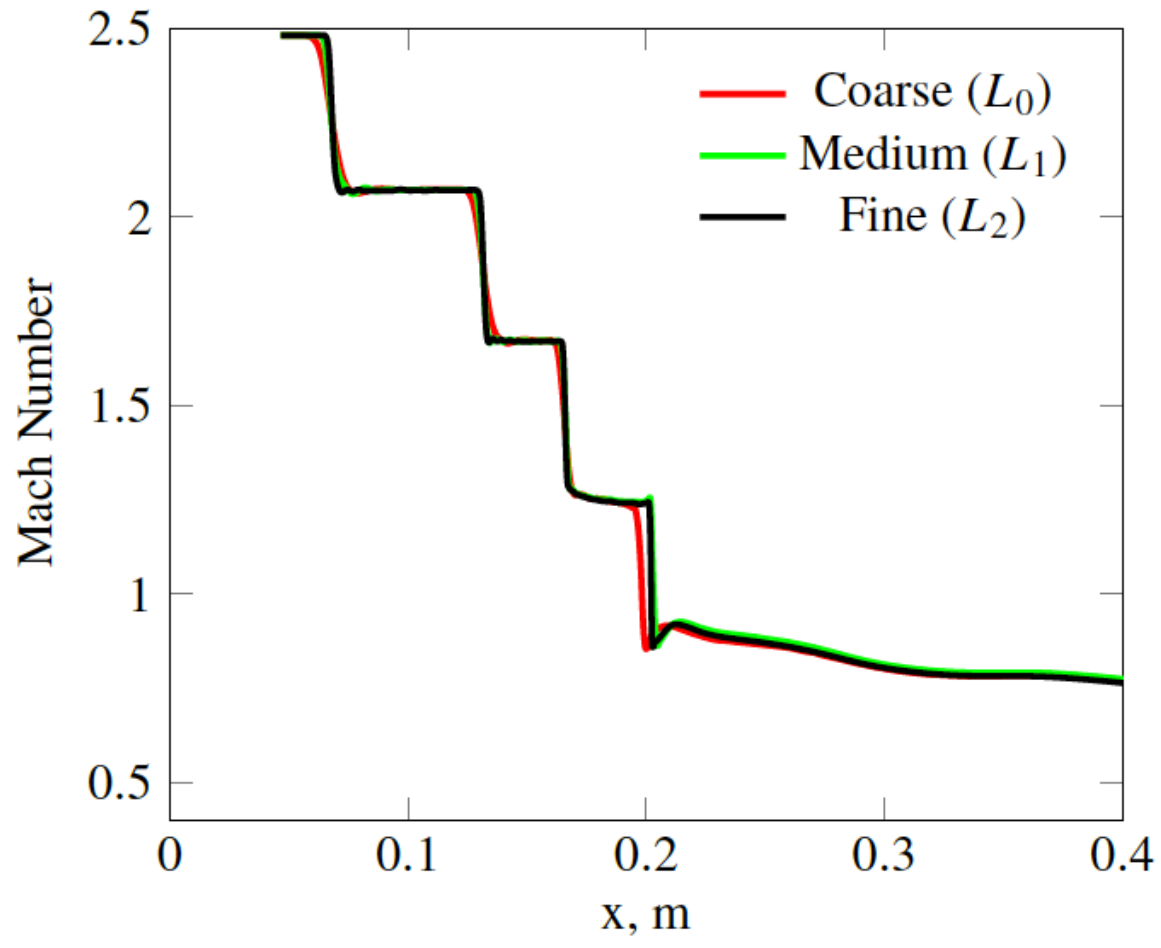

**Fig. 12 Comparison of Mach number along a streamline with different mesh fineness levels.**